\documentclass[aps,nofootinbib,preprintnumbers]{revtex4}
\usepackage{graphicx,amssymb,bm,latexsym}
\newcommand{\bea}{\begin{eqnarray}}
\newcommand{\ena}{\end{eqnarray}}

\renewcommand{\a}{\alpha}
\renewcommand{\b}{\beta}

\newcommand{\p}[1]{(\ref{#1})}

\begin{document}

\preprint{CAS-KITP/ITP-026, KU-TP 018}
\title{Entropy Function and Universality of Entropy-Area
Relation for Small Black Holes}

\author{Rong-Gen Cai$^1$} \email{cairg@itp.ac.cn}

\author{Chiang-Mei Chen$^2$} \email{cmchen@phy.ncu.edu.tw}

\author{Kei-ichi Maeda$^{3,4}$} \email{maeda@waseda.jp}

\author{Nobuyoshi Ohta$^5$} \email{ohtan@phys.kindai.ac.jp}

\author{Da-Wei Pang$^{1,6}$} \email{pangdw@itp.ac.cn}

\affiliation{$^1$Institute of Theoretical Physics, Chinese Academy
of Sciences, P.O.Box 2735, Beijing 100080, China}

\affiliation{$^2$Department of Physics and Center for Mathematics
and Theoretical Physics, National Central University, Chungli 320,
Taiwan}

\affiliation{$^3$Department of Physics, Waseda University, Okubo
3-4-1, Shinjuku, Tokyo 169-8555, Japan}

\affiliation{$^4$Advanced Research Institute for Science and
Engineering, Waseda University, Okubo 3-4-1, Shinjuku, Tokyo
169-8555, Japan}

\affiliation{$^5$Department of Physics, Kinki University,
Higashi-Osaka, Osaka 577-8502, Japan}

\affiliation{$^6$Graduate University of the Chinese Academy of
Sciences, YuQuan Road 19A, Beijing 100049, China}

\date{\today}

\begin{abstract}
We discuss the entropy-area relation for the small black holes with
higher curvature corrections by using the entropy function formalism
and field redefinition method. We show that the entropy $S_{BH}$ of
small black hole is proportional to its horizon area $A$. In
particular we find a universal result that $S_{BH}=A/2G$, the ratio
is two times of Bekenstein-Hawking entropy-area formula in many
cases of physical interest. In four dimensions, the universal
relation is always true irrespective of the coefficients of the
higher-order terms if the dilaton couplings are the same, which is
the case for string effective theory, while in five dimensions, the
relation again holds irrespective of the overall coefficient if the
higher-order corrections are in the GB combination. We also discuss
how this result generalizes to known physically interesting cases
with Lovelock correction terms in various dimensions, and possible
implications of the universal relation.
\end{abstract}

%% REVTEX4
%\pacs{04.20.Jb, 04.65.+e, 98.80.-k}

\maketitle

%%%%%%%%%%%%%%%%%%%%%%%%%%%%%%%%%%%%%%%%%%%%%%%%%%%%%%%%%%%%%%%%%%%%%%
\section{Introduction}
%%%%%%%%%%%%%%%%%%%%%%%%%%%%%%%%%%%%%%%%%%%%%%%%%%%%%%%%%%%%%%%%%%%%%%
Black hole is a fascinating object in gravity which opens a window
to shed light to some quantum effects of gravity. In particular,
black hole behaves like a thermal system which can be described by
macroscopic quantities such as temperature and entropy. The
existence of Hawking temperature indicates that the black holes emit
thermal radiation due to the quantum effect, just like the usual
thermodynamic objects. In Einstein's general relativity, the black
hole entropy is given by quarter of the area of the event horizon,
$S_{BH} = A / 4G$, known as the Bekenstein-Hawking entropy-area
formula,  which inspires the proposal of holographic principle of
gravity~\cite{'t Hooft:1990fr, Susskind:1993ws}.

Extremal charged black holes with degenerate horizon are
particularly simple and crucial in studying many aspects of gravity.
In string theory, extremal black holes are some configurations which
generally preserve partial supersymmetry. For the thermal property,
extremal black holes have zero temperature and therefore are
thermodynamically stable. For some extremal black holes, generically
with the dilaton filed, the degenerate horizon shrinks to a point
(thus the horizon area vanishes) and the singularity is not
protected by a regular horizon to asymptotic observer. This type of
classical solutions is called ``small black hole''
\cite{Sen:2004dp}. If we naively apply the Bekenstein-Hawking
entropy-area formula, the entropy of small black holes vanishes and
the expected quantum degrees of freedom seem to totally
``disappear''. This discrepancy comes from the fact that the general
relativity is only a classical effective theory of quantum gravity.
It has been pointed out that for such kind of black holes, higher
curvature corrections inspired by the low-energy effective action of
quantum theory of gravity, such as Gauss-Bonnet (GB) and Lovelock
terms etc, are expected to stretch the horizon and reproduce correct
entropy corresponding to the microstate degrees of
freedom~\cite{Sen:2004dp}. In this case, the Bekenstein-Hawking
entropy-area relation, $S_{BH}=A/4G$, breaks and should be
significantly revised by the higher curvature terms and could be
obtained by Wald's entropy formula
\cite{Wald:1993nt,Jacobson:1993vj,Jacobson:1994qe}. The near-horizon
geometry of small black holes is $AdS_2 \times S^{D-2}$ after
``stringy cloaking'' and, for this kind of geometry, recently Sen
has developed an elegant approach, the so-called entropy function
formalism,  to calculate the entropy coming from higher curvature
corrections \cite{Sen:2005wa,Sen:2005iz}. (See \cite{Sen:2007qy} for
a review on this topic.)

Following Sen's entropy function approach, various extremal black
holes were investigated, including the two-charge small black holes
from heterotic string compactified on $S^1 \times T^{9-D}$ with
momentum $n$ and winding $w$ on $S^1$ in which the statistical
entropy can be explicitly computed as $S_{\rm statistic} = 4 \pi
\sqrt{nw}$
\cite{Dabholkar:2004yr,Dabholkar:2005by,Dabholkar:2005dt,Dabholkar:2004dq,HMR,Cardoso}.
Among numerous investigation of
small black holes, there is an interesting general property first
observed in \cite{Dabholkar:2004dq} and later in \cite{HMR,
Bak:2005mt, Chen:2006ge}: the black hole entropy and the area of
stretched horizon are related as
\begin{equation}\label{universal}
S_{BH} = \frac1{2G} A.
\end{equation}
This adjusted relation indicates that the higher curvature terms
contribute equal amount of entropy as Hilbert-Einstein action
(scalar curvature). This is an important observation and it is
natural to ask how general or universal this revised relation
(\ref{universal}) is between the entropy and horizon area and what
the physical implications behind the relation are, if any. This is
the basic motivation of this paper.

As a simple example, in Sec.~\ref{d4} we first investigate the
single-charge extremal black hole in four-dimensional dilaton
gravity with general quadratic curvature corrections. We find that
the relation (\ref{universal}) is indeed universal; it is always
true irrespective of the coefficients of the higher-order terms if
the dilaton couplings are the same, which is the case for string
effective theory. In Sec.~\ref{Arbitrary} we generalize the analysis to
arbitrary dimensions and find a constraint on the coefficients of
Ricci and Riemann square terms for the universal
ratio~\p{universal}. In five dimensions, we find that the relation
is again universal irrespective of the overall coefficient if the
higher-order corrections are in the GB combination. This may be
interpreted as another evidence why the higher-order corrections in
string theory should be in this GB combination, in addition to the
known argument of no-ghost condition~\cite{Zwiebach:1985uq}.
These results obtained in Secs.~\ref{d4} and \ref{Arbitrary} are reproduced
by using the field redefinition method in Appendix~\ref{AppendixA},
and there we also show why the entropy of small black hole with
near-horizon geometry $AdS_2\times S^{D-2}$ is always proportional
to its horizon area.

We then generalize our study to the Lovelock gravity with one gauge
field in Sec.~\ref{Lovelock}. In four and five dimensions, there are only GB
terms with common dilaton couplings, and we are uniquely lead to the
universal relation~\p{universal}. It is then natural to examine how
this result may be extended to higher dimensions with more Lovelock terms.
It turns out that this demands relations between the coefficients for
higher-order terms, as given in~\cite{Prester:2005qs}.
We also discuss more general solutions in this type of theories.
Two-charge case is discussed in Appendix~\ref{AppendixB}.

It is worth pointing out that for the particular classes of black
holes in which the microstate counting is available, we find that
the macroscopic entropy satisfies the universal ratio
(\ref{universal})~\cite{Prester:2005qs}.\footnote{In
Ref.~\cite{Prester:2005qs}, the author chooses a special set of
coefficients of Lovelock terms so that the gravity entropy of the
black hole gives the microstate degrees of freedom in any dimension.
In Appendix~\ref{AppendixB} we give a simple proof that with that set of
coefficients, one has $S_{BH}=A/2G$.}
This suggests that the ratio~(\ref{universal}) may be an important
physical principle. How about the inverse statement: can the requirement
that entropy-area ratio~(\ref{universal}) be universal lead to the
macroscopic entropy matching the statistical entropy?
In Sec.~\ref{Microstate}, we examine this issue by looking at
two-charge extremal black holes with only quadratic curvature corrections.
If we assume the ratio (\ref{universal}) as a priori requirement,
then the corresponding macroscopic entropy almost matches the degrees
of freedom from microstate counting up to an overall numerical factor.
It is interesting that this normalization factor is universal independent
of spacetime dimensions. However our result so far is based only on one example.
It would be interesting to find more evidence for this conjecture.
Sec.~\ref{concl} is devoted to conclusion and discussion.

%%%%%%%%%%%%%%%%%%%%%%%%%%%%%%%%%%%%%%%%%%%%%%%%%%%%%%%%%%%%%%%%%%%%%%
\section{Universality in $D=4$ Dilaton gravity with quadratic
curvature terms} \label{d4}
%%%%%%%%%%%%%%%%%%%%%%%%%%%%%%%%%%%%%%%%%%%%%%%%%%%%%%%%%%%%%%%%%%%%%%
As a first example, let us consider the four-dimensional dilatonic
Einstein-Maxwell gravity including most general quadratic curvature
corrections, scalar curvature square, Ricci tensor square and
Riemann tensor square, with arbitrary coefficients $a, b, c$ and
dilaton couplings $\alpha_1, \alpha_2$ and $\alpha_3$. The action
reads
\begin{equation}
I = \frac1{16 \pi G} \int d^4x \sqrt{-g} \left[ R - \frac12
(\partial \phi)^2 - \frac14 \mathrm{e}^{\alpha \phi} F^2 + a
\mathrm{e}^{\alpha_1 \phi} R^2 - b \mathrm{e}^{\alpha_2 \phi}
R_{AB}^2 + c \mathrm{e}^{\alpha_3 \phi} R_{ABCD}^2 \right].
\label{action}
\end{equation}
We should note that $\alpha = \alpha_1 = \alpha_2 = \alpha_3$ is
valid for the low-energy effective theories of heterotic strings.

The near-horizon geometry of a small black hole is supposed to have
$AdS_2 \times S^2$ geometry with constant dilaton and gauge fields
(only electric here)
\begin{equation}
ds^2 = v_1 \left( - r^2 dt^2 + \frac{dr^2}{r^2} \right) + v_2
d\Omega_2^2, \qquad F_{tr} = e, \qquad \phi = \phi_s.
\label{metric}
\end{equation}
There are four parameters characterizing this near-horizon
solution\footnote{This is only a local solution near horizon and its
regular extension globally to asymptotic flat infinity is not always
guaranteed. See the examples discussed in \cite{Chen:2006ge}.}:
$v_1$ and $v_2$ are squares of radii of $AdS_2$ and $S^2$, $e$ is
the gauge field strength and $\phi_s$ is the magnitude of dilaton
filed. The measure of volume element is $\sqrt{-g} = v_1 v_2
\sin\theta$. The following geometrical and physical quantities can
be computed straightforwardly. The Riemann curvature tensors for
$AdS_2$ and $S^2$ are
\begin{equation}
R_{\alpha\beta\gamma\delta} = - \frac1{v_1} ( g_{\alpha\gamma}
g_{\beta\delta} - g_{\alpha\delta} g_{\beta\gamma} ), \qquad
R_{mnpq} = \frac1{v_2} ( g_{mp} g_{nq} - g_{mq} g_{np}),
\end{equation}
and the Ricci tensor and scalar curvature are
\begin{equation}
R_{\alpha\beta} = - \frac1{v_1} g_{\alpha\beta}, \qquad R_{mn}=
\frac1{v_2} g_{mn},
\qquad R = - \frac2{v_1} + \frac2{v_2}.
\end{equation}
The associated quadratic combinations of curvature and the field
strength square $F^2$ are
\begin{equation}
R_{AB}^2 = \frac2{v_1^2} + \frac2{v_2^2}, \qquad
R_{ABCD}^2 = \frac4{v_1^2} + \frac4{v_2^2}, \qquad F^2 = - \frac{2 e^2}{v_1^2}.
\end{equation}
Here we have used the notation that the upper case indices $A,B, \ldots$ run
over the whole spacetime, while $\a,\b,\ldots$ over the AdS space and
$m,n,\ldots$ over the sphere. This notation is used throughout this paper.

The entropy of small black holes can be derived from the entropy
function $f$ defined by \cite{Sen:2007qy}
\begin{equation}
I = \int dt dr f,
\end{equation}
which is basically the integration of the Lagrangian over the
spherical part of geometry. The explicit form of the entropy
function in this particular theory is
\begin{equation}
f \equiv \frac{v_1 v_2}{4 G} \left[ \frac2{v_2} - \frac2{v_1} +
\frac12 \mathrm{e}^{\alpha \phi_s} \frac{e^2}{v_1^2} + 4 a
\mathrm{e}^{\alpha_1 \phi_s} \left( \frac1{v_1} - \frac1{v_2}
\right)^2 - \left( 2 b \mathrm{e}^{\alpha_2 \phi_s} - 4 c
\mathrm{e}^{\alpha_3 \phi_s} \right) \left( \frac1{v_1^2} +
\frac1{v_2^2} \right) \right].
\end{equation}
The physical charge $q$ corresponding to the field strength $e$ at
the horizon is given by \cite{Sen:2007qy}
\begin{equation}
q = \frac{\partial f}{\partial e} = \frac{\mathrm{e}^{\alpha
\phi_s}}{4G} \; \frac{v_2}{v_1} \; e.
\end{equation}
The values of $v_1, v_2$ and $\phi_s$ are determined by extremizing
the entropy function, i.e. by the following equations
\begin{eqnarray}
G \frac{\partial f}{\partial v_1} &=& \frac12 - \frac{e^2 v_2}{8
v_1^2} \mathrm{e}^{\alpha \phi_s} - \left( \frac{v_2}{v_1^2} -
\frac1{v_2} \right) \left( a \mathrm{e}^{\alpha_1 \phi_s} -
\frac{b}2 \mathrm{e}^{\alpha_2 \phi_s} + c \mathrm{e}^{\alpha_3
\phi_s} \right) = 0,
\nonumber\\
\label{deq}
G \frac{\partial f}{\partial v_2} &=& - \frac12 + \frac{e^2}{8 v_1}
\mathrm{e}^{\alpha \phi_s} - \left( \frac{v_1}{v_2^2} - \frac1{v_1}
\right) \left( a \mathrm{e}^{\alpha_1 \phi_s} - \frac{b}2
\mathrm{e}^{\alpha_2 \phi_s} + c \mathrm{e}^{\alpha_3 \phi_s}
\right) = 0,
\\
G \frac{\partial f}{\partial \phi_s} &=& v_1 v_2 \left[ \frac{\alpha
e^2}{8 v_1^2} \mathrm{e}^{\alpha \phi_s} + a \alpha_1
\mathrm{e}^{\alpha_1 \phi_s} \left( \frac1{v_1} - \frac1{v_2}
\right)^2 - \left( \frac{b \alpha_2}2 \mathrm{e}^{\alpha_2 \phi_s} -
c \alpha_3 \mathrm{e}^{\alpha_3 \phi_s}  \right) \left(
\frac1{v_1^2} + \frac1{v_2^2} \right) \right] = 0, \nonumber
\end{eqnarray}
and the black hole entropy can be obtained from the entropy function
by a Legendre transformation with respect to the physical charge:
\begin{equation}
S_{BH} = 2 \pi (e q - f).
\end{equation}
The solution to the set of equations~\p{deq} is
\begin{equation}
v_1 = v_2 = 4 G^2 q^2 \mathrm{e}^{- \alpha \phi_s}.
\label{v}
\end{equation}
The remaining part of the solution depends on the values of
parameters in the Lagrangian and there are three cases to be
distinguished.
\begin{enumerate}
\item $\a_2 \neq 0$: \\
In this case, we can solve the last equation in (\ref{deq}) as
\begin{equation}
b = \frac{2 G^2 \alpha q^2}{\alpha_2} \mathrm{e}^{- (\alpha +
\alpha_2) \phi_s} + \frac{2 c \alpha_3}{\alpha_2}
\mathrm{e}^{(\alpha_3 - \alpha_2) \phi_s},
\label{dil}
\end{equation}
which should be understood as an equation for determining $\phi_s$ as a function
of the physical charge $q$
but it is generally difficult to give its explicit form. The area is given by
\begin{equation}
A = 4 \pi v_2 = 16 \pi G^2 q^2 \mathrm{e}^{-\alpha \phi_s},
\label{area}
\end{equation}
and the entropy is
\begin{equation}
S_{BH} = 2 \pi (e q - f) = \frac{4 \pi}{G} \left[ \frac{G^2 (\alpha
+ \alpha_2) q^2}{\alpha_2} \mathrm{e}^{-\alpha \phi_s} + \frac{c
(\alpha_3 - \alpha_2)}{\alpha_2} \mathrm{e}^{\alpha_3 \phi_s}
\right].
\end{equation}
Therefore, the entropy-area relation is
\begin{equation}
S_{BH} = \frac1{4 G} \left[ \frac{\alpha + \alpha_2}{\alpha_2} +
\frac{c (\alpha_3 - \alpha_2)}{G^2 \alpha_2 q^2} \mathrm{e}^{(\alpha
+ \alpha_3) \phi_s} \right] A.
\label{ratio}
\end{equation}
For fixed coefficients $a, b, c$ and exponents $\a, \a_1,\a_2$ and $\a_3$,
$\phi_s$ is a function of the physical charge $q$.
It can be checked by using Eq.~\p{dil} that the square bracket in Eq.~\p{ratio}
is constant only for $\a_2=\a_3$. Then the entropy is always proportional
to the area {\it irrespective of the coefficients $a, b, c$ and exponents
$\a$'s}, a manifestation of holographic principle. Note that the
relation~(\ref{ratio}) is independent of the value of $\alpha_1$.
This is because the contribution of the scalar curvature vanishes in
the solution.

The fact that the entropy is proportional to the area may appear somewhat
trivial for the spherically symmetric small black holes.
However we emphasize that Eq.~\p{ratio} does not mean the proportionality.
Given the entropy and the area expressed in terms of solutions, we can formally
write their relation like Eq.~\p{ratio}, but it has to be checked
whether the coefficient is a constant or not. Indeed, in our above
example, we have shown that we must have $\a_2=\a_3$; otherwise the coeffcient
changes when we change the physical charge. We cannot say that they are
proportional for such a general case.

In the string effective theory, $\alpha = \alpha_2 = \alpha_3$ and
the relation (\ref{ratio}) reduces to the universal one
(\ref{universal}). The relation~\p{universal} is universally true
irrespective of the precise values of the coefficients $a, b, c$ and
exponents $\a$'s, and only the relative magnitudes of the exponents
$\a$'s are important. In particular, it is not necessary that the
higher-order corrections are in the GB combination. Thus the dilaton
coupling appears to automatically adjust these values to produce the
universal result~\p{universal} in string theories. Here we can solve
Eq.~(\ref{dil}) for $\phi_s$ to obtain
\begin{equation}
\phi_s = \frac{1}{2 \alpha} \ln\left( \frac{2 G^2 q^2}{b - 2c}
\right).
\end{equation}

\item $\alpha_3 \neq 0$: \\
We have basically the same equations as in item 1 with $(\alpha_2,
b)$ and $(\alpha_3, -2c)$ interchanged, and get the universal
relation~\p{universal} for $\alpha = \alpha_2 =\alpha_3$.

\item $\alpha_2 = \alpha_3 = 0$: \\
In this case, the last equation of (\ref{deq}) gives constraint
$v_1 v_2 \frac{\alpha e^2}{8 v_1^2} \mathrm{e}^{\alpha \phi_s} = 0,$ namely
\begin{equation}
\qquad \alpha v_2 = 0.
\end{equation}
For the case $v_2 = 0$ and $\a \neq 0$, which amounts to
$\phi_s \to \alpha \times \infty$ from \p{v} (the exceptional case is neutral
solution $q = 0$ which is not of our interest), both area and entropy vanish.
The near-horizon geometry~\p{metric} is not valid in this case, and we expect that
further higher curvature corrections are necessary.

For the other case $\alpha = 0$, the dilaton is completely decoupled and
the system is not of our interest. Nevertheless, let us see what we get.
The horizon area and the entropy are given by (this is the only case in
which $f \ne 0$)
\begin{equation}
A = 4 \pi v_2 = 16 \pi G^2 q^2, \qquad\qquad S_{BH} = 2 \pi (e q - f)
= 4 \pi G q^2 + \frac{2 \pi (b - 2c)}{G},
\end{equation}
independently of the value of $\alpha_1$. We find the entropy-area
relation
\begin{equation}
S_{BH} = \frac1{4G} \left( 1 + \frac{b - 2c}{2 G^2 q^2} \right) A.
\label{lbh1}
\end{equation}

When the higher-order corrections are absent, this gives
\begin{equation}
S_{BH} = \frac1{4G} A, \label{lbh2}
\end{equation}
which is nothing but the result for Einstein gravity.
\end{enumerate}

Because the scalar curvature vanishes in the solution, their
corrections do not play any role in the above evaluation. So the
above results are always valid even if other higher-order
corrections in the scalar curvature are added to the action~(\ref{action})
\begin{equation}
\sum_{n=2} c_n \mathrm{e}^{\alpha_n \phi} R^n.
\end{equation}

We have thus found that small black holes in four dimensions are
very special in that they have the universal
relation~(\ref{universal}) for arbitrary combination of the
curvature square terms, not just for the GB combination as usually
supposed to be. We will see that this is no longer true in higher
dimensions with higher curvature corrections. However if we consider
Lovelock type corrections, we get the relation~(\ref{universal}).

%%%%%%%%%%%%%%%%%%%%%%%%%%%%%%%%%%%%%%%%%%%%%%%%%%%%%%%%%%%%%%%%%%%%%%
\section{Arbitrary Dimensions} \label{Arbitrary}
%%%%%%%%%%%%%%%%%%%%%%%%%%%%%%%%%%%%%%%%%%%%%%%%%%%%%%%%%%%%%%%%%%%%%%
In this section, we examine the entropy-area relation
 for the general theory in arbitrary $D$ dimensions with the action
\begin{equation}
I = \frac1{16 \pi G} \int d^Dx \sqrt{-g} \left[ R - \frac12
(\partial \phi)^2 - \frac14 \mathrm{e}^{\alpha \phi} F^2 + a
\mathrm{e}^{\alpha_1 \phi} R^2 - b \mathrm{e}^{\alpha_2 \phi}
R_{AB}^2 + c \mathrm{e}^{\alpha_3 \phi} R_{ABCD}^2 \right].
\label{action_D}
\end{equation}
 The extremal black
holes in $D$ dimensions are assumed to have near-horizon
geometry $AdS_2 \times S^{D-2}$ and the relevant geometric
quantities are
\begin{eqnarray}
R_{\alpha\beta} = - \frac1{v_1} g_{\alpha\beta}, \quad
R_{mn}= \frac{D-3}{v_2} g_{mn}, \quad R = - \frac2{v_1} + \frac{(D-2)
(D-3)}{v_2}, \label{Ricci}
\\
R_{AB}^2 = \frac2{v_1^2} + \frac{(D-2)(D-3)^2}{v_2^2}, \qquad
R_{ABCD}^2 = \frac4{v_1^2} + \frac{2 (D-2) (D-3)}{v_2^2}, \qquad F^2
= - \frac{2 e^2}{v_1^2}.
\label{Ricci_squared}
\end{eqnarray}
It is straightforward to compute the entropy function
\begin{eqnarray}
f &\equiv& \frac{\Omega_{D-2} v_1 v_2^{\frac{D-2}2}}{16 \pi G}
\Biggl[ \frac{(D-2)(D-3)}{v_2} - \frac2{v_1} + \frac12
\mathrm{e}^{\alpha \phi_s} \frac{e^2}{v_1^2} + a
\mathrm{e}^{\alpha_1 \phi_s} \left( \frac2{v_1} - \frac{(D-2)
(D-3)}{v_2} \right)^2
\nonumber\\
&& \qquad - b \mathrm{e}^{\alpha_2 \phi_s} \left( \frac2{v_1^2} +
\frac{(D-2) (D-3)^2}{v_2^2} \right) + 2 c \mathrm{e}^{\alpha_3
\phi_s} \left( \frac2{v_1^2} + \frac{(D-2) (D-3)}{v_2^2} \right)
\Biggr],
\end{eqnarray}
where $\Omega_{D-2} = 2 \pi^{\frac{D-1}2}/\Gamma(\frac{D-1}2)$ is
surface area of unit sphere $S^{D-2}$. The relation of physical
charge $q$ and the field strength $e$ is
\begin{equation}
e = \frac{16 \pi G v_1}{\Omega_{D-2} \, \mathrm{e}^{\alpha \phi_s}
\, v_2^{\frac{D-2}2}} \, q.
\end{equation}

In general, it is complicated to find the extremal value of $f$ for
arbitrary dilaton couplings. For simplicity we focus on the special
case of string effective theory in which $\alpha_1 = \alpha_2 =
\alpha_3= \alpha$ (all dilaton couplings are equal). The conditions
for $f$ to have extremal value give the relation
\begin{equation}
v_2 = \frac12 (D-2) (D-3) v_1, \label{v1_v2}
\end{equation}
which implies vanishing scalar curvature, $R = 0$, and the constants
$v_1$ and $\phi_s$ must satisfy the equations
\begin{eqnarray}
v_1 &=& \frac{(D-3) e^2 + 8 c (D-4)}{D (D-3)} \, \mathrm{e}^{\alpha
\phi_s}, \label{v_1}
\\
b &=& \frac14 \frac{(D-2) (D-3) e^2 + 8 c (D^2 - 5D + 8)}{D (D-3)}.
\label{b}
\end{eqnarray}
Note that $\partial f / \partial \phi_s = 0$ implies either $\alpha
= 0$ in which dilaton is totally decoupled (and we are not
interested in this) or $f = 0$. The explicit solutions for $v_1$ and
$\phi_s$ are complicated. However, the explicit forms are not
essential for deriving the entropy-area ratio.

The area of event horizon is
\begin{equation}
A = \Omega_{D-2} v_2^{\frac{D-2}2} = \Omega_{D-2} \left[ (2 D b - 6b
- 4c) \, \mathrm{e}^{\alpha \phi_s} \right]^{\frac{D-2}2},
\end{equation}
and the entropy is ($f = 0$)
\begin{equation}
S_{BH} = 2 \pi (e q - f) = \frac{\Omega_{D-2} v_2^{\frac{D-2}2} \,
\left[ D (D-3) b - 2 (D^2 - 5D + 8) c \right]}{8 G (D b - 3b - 2c)}.
\label{BH_entropy_any_dimension}
\end{equation}
Thus the entropy-area relation is
\begin{equation}\label{SAanyD}
S_{BH} = \frac{D (D-3) b - 2 (D^2 - 5D + 8) c}{8 G (D b - 3b - 2c)} A.
\end{equation}
We thus again find that the entropy is always proportional to the
area irrespective of the coefficients $a, b, c$ and exponents
$\a$'s. We can derive this relation from the Bekenstein-Hawking
formula by field redefinition (see Appendix \ref{AppendixA}).

Let us also check when we get the relation $S_{BH} = A/2G$. It turns
out that we must have either $D = 4$ or
\begin{equation}\label{Rbc}
(D - 3) b = 2 (D - 1) c.
\end{equation}
This means that the corrections are in the GB combination for $D =
5$. In this case, only the relative magnitudes of the coefficient
are determined and the universal relation~\p{universal} is true
irrespective of the overall factor. If we regard the universal
relation as a physically important principle, this might be another
evidence why the higher-order corrections must come in the GB
combination in string effective theories in addition to the
ghost-free condition~\cite{Zwiebach:1985uq}.

For theories in $D > 5$, \p{Rbc} does not give the GB combination.
Does this mean that the relation~\p{universal} is not valid in
dimensions higher than five? Considering that GB term is the leading
correction in heterotic string theory, this problem gets physical
interest. Rather than hastily jumping to such a conclusion, we
suggest that this is an indication that we should consider more
higher curvature corrections, such as the Lovelock terms. We are now
going to consider this possibility and see that indeed we can obtain the
relation~\p{universal} in higher dimensions.

%%%%%%%%%%%%%%%%%%%%%%%%%%%%%%%%%%%%%%%%%%%%%%%%%%%%%%%%%%%%%%%%%%%%%%
\section{Lovelock Gravity}\label{Lovelock}
%%%%%%%%%%%%%%%%%%%%%%%%%%%%%%%%%%%%%%%%%%%%%%%%%%%%%%%%%%%%%%%%%%%%%%
In this section, let us consider the dilatonic Einstein-Maxwell
theory with Lovelock higher curvature corrections. In the string frame,
the action is
\begin{equation}
I = \frac1{16\pi G} \int d^Dx \sqrt{-g} S \, \sum_{m=1}
\alpha'^{m-1} \mathcal{L}_m,
\end{equation}
where $S$ on the rhs is the dilaton field. The leading term in the Lagrangian is
\begin{equation}
\mathcal{L}_1 = R - \frac12 (\partial \phi)^2 - \frac14 F_{[2]}^2,
\end{equation}
and the higher-order terms are
\begin{equation}
\mathcal{L}_m = \frac{\lambda_m}{2^m} \; \delta^{\rho_1 \sigma_1
\cdots \rho_m \sigma_m}_{\mu_1 \nu_1 \cdots \mu_m \nu_m} \; R^{\mu_1
\nu_1}{}_{\rho_1 \sigma_1} \cdots R^{\mu_m \nu_m}{}_{\rho_m
\sigma_m}, \qquad m = 2, \cdots, [D/2],
\end{equation}
where $\lambda_m$ are dimensionless parameters. This action, written
in the string frame, is a generalization of those actions considered
in the previous sections to higher dimensions with higher curvature
terms for equal dilaton coupling.

The entropy function for the black hole with $AdS_2 \times S^{D-2}$
near-horizon geometry is~\cite{Prester:2005qs}
\begin{equation}\label{Lovelock_f}
f = \frac{\Omega_{D-2} u_S v_1 v_2^{\frac{D-2}2}}{16 \pi G} \left\{
\frac{e^2}{2 v_1^2} + \sum^{[D/2]}_{m=1} \frac{(D-2)!}{(D-2m)!}
\frac{\alpha'^{m-1} \lambda_m}{v_2^m} \left[ (D-2m)(D-2m-1) -
\frac{2 m v_2}{v_1} \right] \right\},
\end{equation}
where $u_S$ is the near-horizon value of dilaton field $S$. Here the
Hilbert-Einstein term (scalar curvature) is included as the $m = 1$
term by defining $\lambda_1 = 1$. The summation in the entropy
function (\ref{Lovelock_f}) can be rearranged in terms of power of
$v_2$ as
\begin{equation}
\sum^{[D/2]}_{m=1} \Bigl( \cdots \Bigr) = \sum^{[D/2]}_{m=1} K_m \,
v_2^{1 - m},
\end{equation}
where the coefficients $K_m$, functions of $v_1$, are
\begin{equation}
K_1 = - \frac2{v_1}, \qquad K_m = \alpha'^{m - 2} \frac{(D - 2)!}{(D
- 2m)!} \left( \lambda_{m-1} - \frac{2 m \alpha'}{v_1} \lambda_m
\right), \quad m \ge 2.
\end{equation}
The equation $\partial f/\partial u_S = 0$ for extremal value of
$f$ requires $f = 0$,
\begin{equation}
\label{Equs}
\sum^{[D/2]}_{m=2} K_m \, v_2^{1 - m} + \frac{e^2}{2 v_1^2} -
\frac2{v_1} = 0,
\end{equation}
and the equation $\partial f/\partial v_2 = 0$ gives
\begin{equation}
\label{Eqv2}
\sum^{[D/2]}_{m=2} (m - 1) K_m \, v_2^{- m} = 0.
\end{equation}
The last equation $\partial f/\partial v_1 = 0$ is more complicated
\begin{equation}\label{Eqv1}
\sum^{[D/2]}_{m=2} \frac{(D - 2)!}{(D - 2m)!} m \alpha'^{m-1}
\lambda_m \; v_2^{1-m} - \frac{e^2}{2 v_1} + 1 = 0.
\end{equation}
Finally, the value of $u_S$ is determined from the physical charge
defined by
\begin{equation}\label{Eqe}
q = \frac{\partial f}{\partial e} = \frac{\Omega_{D-2} u_S
v_2^{\frac{D-2}2}}{16 \pi G v_1} \; e.
\end{equation}

Now our goal is to solve equations (\ref{Equs})--(\ref{Eqe}) to
find the values of $u_S, v_1, v_2$ and $e$ in terms of the
physical charge $q$ and the parameters $\lambda_m$'s in the
theory. Formally the horizon area and the black hole entropy
($f=0$) can be expressed as
\begin{equation}
\label{LovelockAS}
A = \Omega_{D-2} u_S v_2^{\frac{D-2}2} = 16 \pi G \; \frac{v_1
q}{e}, \qquad S_{BH} = 2 \pi (e q - f) = 2 \pi \, e q,
\end{equation}
and the ratio is
\begin{equation}
\label{LovelockSAratio}
S_{BH} = \frac{e^2}{8 G v_1} A.
\end{equation}
Here again the entropy is proportional to the area. Note that here
the universal relation~\p{universal} is equivalent to
\begin{equation}
v_1 = \frac{e^2}{4}. \label{v1}
\end{equation}

\subsection{Universal relation in $D=4,5$ and generalization to
higher dimensions}
%%%%%%%%%%%%%%%%%%%%%%%%%%%%%%%%%%%%%%%%%%%%%%%%%%%%%%%%%%%%%%%%%%%%%%

In $D=4,5$, there are only GB terms. Let us first consider $D=4$
though this is a special case of those discussed in Sec.~\ref{d4}.
{}From Eqs.~\p{Eqv2} and \p{Equs}, we find\footnote{Note that in
this case, the unique solution of (\ref{Eqv2}) is $K_2=0$.}
\begin{equation}
v_1 = 4 \alpha' \lambda_2, \qquad e^2 = 4 v_1 = 16 \alpha'
\lambda_2, \label{v1v2}
\end{equation}
which already assures that the universal relation~\p{universal}
holds in this case (see (\ref{LovelockSAratio})). The remaining
Eq.~\p{Eqv1} then gives $v_2$ as
\begin{equation}
v_2 = 4 \alpha' \lambda_2 = v_1. \label{e}
\end{equation}
The dilaton field can be determined as $u_s=Gq/(4\alpha' \lambda_2)$
through (\ref{Eqe}). Thus all moduli fields are given in terms of
the physical charge $q$ and the parameter $\lambda_2$. Note that
here the relation $S=A/2G$ is independent of the value of
$\lambda_2$, namely, $\lambda_2$ can be arbitrary, in agreement with
the result in Sec.~\ref{d4}.

Next for the case $D=5$,  we obtain
\begin{equation}
v_1 = 4 \alpha' \lambda_2, \qquad e^2 = 4 v_1 = 16 \alpha'
\lambda_2,
\end{equation}
and $v_2= 12\alpha'\lambda_2$. Thus, once again, we have the
universal relation~\p{universal} irrespective of the value of
$\lambda_2$, in agreement with the results in Sec.~\ref{Arbitrary}.
Of course, if we  want to match the result with the microstate
counting, the value of $\lambda_2$ should be fixed in some way (see
for example, \cite{Prester:2005qs} or \cite{Gross:1986mw} where
$\lambda_2$ is determined to be $\frac18$). Note that in the case of
$D=5$, one has also $K_2=0$.

When we go up to $D=6$ and $D=7$, the next Lovelock term
contributes. We know from Eq.~\p{LovelockAS} that if one requires
that the entropy should be independent of the spacetime
dimension,\footnote{Similar argument based on the microstate
counting is given in Ref.~\cite{Prester:2005qs}.} $e$ must be
independent of the dimension (note that $q$ is a physical charge,
which is an input quantity). On the other hand, since we already
know that $K_2=0$ from the case of $D=4$ and 5, we conclude that
$K_3=0$ from Eq.~\p{Eqv2}. The only solution of Eq.~\p{Equs} is then
$v_1=4e^2$, which is independent of the spacetime dimension. We thus
come back to the universal result~\p{universal}, again whatever the
value of $\lambda_2$ is. We see that this is the value that new
higher-order correction does not modify the solution for the
lower-dimensional solutions. The condition $K_3=0$ gives
\begin{equation}
\lambda_3 = \frac{v_1}{6 \alpha'} \lambda_2. \label{l3}
\end{equation}
Combined with Eq.~\p{v1v2}, Eq.~\p{l3} leads to
\begin{equation}
\lambda_3 = \frac{1}{3!} (2 \lambda_2)^2,
\end{equation}
so that we can see that only $\lambda_2$ is a free parameter.

As we have observed, the solution $v_2$ depends on dimension $D$,
but $v_1$ does not. It is natural to take the value of $v_1$ as it
is in all dimensions~\cite{Prester:2005qs}. Under this condition,
Eqs.~\p{Equs} and (\ref{Eqv2}) determine the solution\footnote{Note that each
$K_i$ is different by a numerical factor in different dimensions.}
by $K_m = 0$ for $m \ge 2$, which completely fix all other
$\lambda_m$ by $\lambda_2$ via
\begin{equation}
\label{lambda_m}
\lambda_m = \frac{v_1}{2 m \alpha'} \; \lambda_{m-1} = \frac1{m!}
\left( \frac{v_1}{2 \alpha'} \right)^{m-1}
= \frac{1}{m!} (2\lambda_2)^{m-1}.
\end{equation}
where we also used the result~\p{v1v2}. Obviously the universal relation
$S_{BH} = A/2G$ still holds by (\ref{LovelockSAratio}) and \p{v1}.
Note that $v_2$ is determined by Eq.~\p{Eqv1} and it is easy to
show that it has at least one positive solution. This point will
be discussed later in more general context.

The relation~\p{v1v2} also indicates that $e$ is not a free parameter
but its value is fixed by parameters $\alpha'$ and $\lambda_2$ in
the Lagrangian. Thus the value of $e$ does not appear to contain any
information on the magnitude of charge. This seems peculiar since
$e$ is the value of $F_{tr}$ at horizon. This strange result is due
to the fact that the dependence of field strength on charge
parameter is hidden in the string frame by the coupling of the
dilaton. From Eq. (\ref{Eqe}), one can see that the value $u_S$ of
the dilaton at horizon actually carries the charge information. The
relation between the field strength and charge becomes more
transparent if the solution is presented in the Einstein frame.

Finally we note that there may be other possibility for achieving
the relation $S_{BH} = A/2G$ if we do not require that the
coefficients be determined step by step from lower dimensions to
higher dimensions as above. For example, consider $D=10$ with free
parameters $\lambda_2, \lambda_3$ and $\lambda_4$. Let us take
$\lambda_2 = \frac18$ and $\lambda_3 = 0$, which are the known
values for these in string theory~\cite{Gross:1986mw}. To have the
relation $S_{BH} = A/2G$, we must have $v_1 = e^2/4$ and the
extremal conditions of the entropy function give
\begin{equation}
v_1 = \frac{7(19823-31 \sqrt{89761})}{165243} \alpha' > 0, \quad v_2
= \frac{7}{223} (195 + \sqrt{89761}) \alpha' > 0, \quad \lambda_4 =
\frac{49(366563+3737 \sqrt{89761})}{15968976480}.
\end{equation}
Thus the relation~\p{universal} is possible in arbitrary dimensions
by adjusting $\lambda$'s even if we do not require $\lambda$'s are
determined in lower dimensions. These values need not be taken
serious because it is known that there are other corrections
in the order of $R^4$ in the heterotic strings~\cite{Gross:1986mw}.
(The string case in higher dimensions is beyond the scope of this paper
and may be a subject of future study.)
We now discuss this kind of general solutions in more detail.

\subsection{General solution}
%%%%%%%%%%%%%%%%%%%%%%%%%%%%%%%%%%%%%%%%%%%%%%%%%%%%%%%%%%%%%%%%%%%%%%
In this section, we show that we can formally solve
Eqs.~(\ref{Equs})--(\ref{Eqe}) for $u_S, v_1, v_2$ and $e$, and
obtain positive $v_1$ and $v_2$. Eliminating $e^2$, which contains
$u_S$, from Eqs. (\ref{Equs}) and (\ref{Eqv1}), we find
\begin{equation}
\sum^{[D/2]}_{m=2} \frac{(D - 2)!}{(D- 2m)!} \, \alpha'^{m - 2}
\left( \lambda_{m-1}v_1 -  m \alpha' \lambda_m \right) \, v_2^{1-m}
- 1 = 0.
\end{equation}
This equation is rewritten as
\begin{equation}
{\cal A}_1 v_1 = {\cal B}_1, \label{eq_v1v2_1}
\end{equation}
where
\begin{eqnarray}
{\cal A}_1 &=& \sum^{[D/2]}_{m=2} \frac{(D - 2)!}{(D - 2m)!} \,
\alpha'^{m - 2} \lambda_{m-1} \, v_2^{1-m},
\nonumber \\
{\cal B}_1 &=& \sum^{[D/2]}_{m=2} \frac{m \, (D - 2)!}{(D - 2m)!} \,
\alpha'^{m - 1} \lambda_m \, v_2^{1-m} + 1 \,.
\label{b1}
\end{eqnarray}
Now Eq.~(\ref{Eqv1}) has a simple expression
\begin{equation}
e^2 = 2 v_1 {\cal B}_1,
\end{equation}
and Eq.~(\ref{Eqv2}) can also be rewritten in the form of
\begin{equation}
{\cal A}_2 v_1 = {\cal B}_2,
\label{eq_v1v2_2}
\end{equation}
where
\begin{eqnarray}
{\cal A}_2 &=& \sum^{[D/2]}_{m=2} \frac{(m-1) \, (D - 2)!}{(D- 2m)!}
\, \alpha'^{m - 2} \lambda_{m-1} \, v_2^{1-m} \,,
\nonumber\\
{\cal B}_2 &=& 2 \sum^{[D/2]}_{m=2} \frac{m(m-1)\, (D - 2)!}{(D-
2m)!} \, \alpha'^{m - 1} \lambda_m \, v_2^{1-m} \,.
\end{eqnarray}
Eliminating $v_1$ from Eqs. (\ref{eq_v1v2_1}) and (\ref{eq_v1v2_2}),
we find the equation for $v_2$ as
\begin{equation}
{\cal F}(v_2) \equiv v_2^{2[D/2]-2} \left[ {\cal A}_2 {\cal B}_1 -
{\cal A}_1 {\cal B}_2 \right] = 0 \,,
\label{eq_v2}
\end{equation}
whose explicit form is
\begin{eqnarray}
{\cal F}(v_2) = \left( \sum^{[\frac{D}2]}_{m=2} \frac{(m - 1) (D -
2)!}{(D - 2m)!} \alpha'^{m-2} \lambda_{m-1} v_2^{[\frac{D}2]-m}
\right) \left( v_2^{[\frac{D}2]-1} + \sum^{[\frac{D}2]}_{m=2}
\frac{m (D - 2)!}{(D - 2m)!} \alpha'^{m - 1} \lambda_m
v_2^{[\frac{D}2] - m} \right)
\nonumber \\
- 2 \left( \sum^{[\frac{D}2]}_{m=2} \frac{(D - 2)!}{(D - 2m)!}
\alpha'^{m - 2} \lambda_{m-1} v_2^{[\frac{D}2] - m} \right) \left(
\sum^{[\frac{D}2]}_{m=2} \frac{m(m-1) (D - 2)!}{(D - 2m)!}
\alpha'^{m - 1} \lambda_m v_2^{[\frac{D}2] - m} \right).
\end{eqnarray}
The equation ${\cal F}(v_2)=0$ is the $(2[D/2]-3)$ order algebraic
equation for $v_2$. Once we find the solution for $v_2$, we obtain
the solutions for $v_1$ and $u_S$ as
\begin{eqnarray}
v_1 &=& {{\cal B}_1 \over {\cal A}_1} = {{\cal B}_2 \over {\cal
A}_2} \,, \label{Eqv1_BA}
\\
u_S^2 &=& {1 \over 2} \left( {16 \pi G \, q \over \Omega_{D-2}}
\right)^2 {1 \over {\cal A}_1 v_2^{D-2}} \,.
\end{eqnarray}
Hence if there exists a solution of ${\cal F}(v_2) = 0$ for $v_2$,
we obtain the solution for any coupling constant $\lambda_m$ and any
charge $q$.

Here let us give two simple examples.
\begin{enumerate}
\item Case of $[D/2]=2$ ($D=4$ or $5$): \\
The equation for $v_2$ is very simple and the solution is $v_2 = 2
(D-2)(D-3) \alpha' \lambda_2$. We find $v_1 = 4 \alpha'
\lambda_2$, that is $v_2 = (D-2)(D-3)v_1/2$. (Note that
$\lambda_1= 1$.)

\item Case of $[D/2]=3$ ($D=6$ or $7$): \\
${\cal F}(v_2) = 0$ gives the cubic equation for $v_2$ as
\begin{eqnarray}
&& v_2^3 - 4 (2 D - 7) \alpha' \lambda_2 v_2^2 - 9 (D-2) (D-3) (D-4)
(D-5) \alpha'^2 \lambda_3 v_2
\nonumber \\
&& \qquad - 6 (D-2) (D-3) (D-4)^2 (D-5)^2 \alpha'^3 \lambda_2
\lambda_3 = 0 \,.
\end{eqnarray}
\end{enumerate}

\medskip
In general, it is difficult to obtain an explicit solution of $v_2$
from ${\cal F}(v_2) = 0$. Now, for this solution to be acceptable,
we must show that there exists at least one positive solution for
$v_2$. Since ${\cal F}(v_2) = (D-2)(D-3) v_2^{2[D/2]-3} + \cdots$,
${\cal F}(v_2) \to + \infty$ as $v_2 \to \infty$. Also
\begin{equation}
{\cal F}(v_2)|_{v_2=0} = - {[{D \over 2}] \left( [{D \over 2}] - 1
\right) \{ (D-2)! \}^2 \over \{ \left( D - 2 [{D \over 2}]
\right)!\}^2 } \alpha'^{2 [{D \over 2}] - 3} \lambda_{[{D \over 2}]
- 1} \lambda_{[{D \over 2}]} \, < 0 \,,
\end{equation}
if $\lambda_{[{D \over 2}] - 1} \lambda_{[{D \over 2}]} > 0$.
Because the continuous function ${\cal F}(v_2)$ changes from
negative value at $v_2=0$ to infinity at $v_2=\infty$, we have at
least one positive solution for $v_2$ when this last condition is
satisfied.

As for the entropy, inserting our solution into the definition $S =
2 \pi e q$, we have
\begin{equation}
S_{BH} = {\cal B}_1 \times {A \over 4G}.
\end{equation}
We can reproduce our earlier results using these formulae. We see
from Eq.~\p{b1} that ${\cal B}_1>1$ if $\lambda_m$'s are all
positive.

%%%%%%%%%%%%%%%%%%%%%%%%%%%%%%%%%%%%%%%%%%%%%%%%%%%%%%%%%%%%%%%%%%%%%%
\section{Matching microstate counting}\label{Microstate}
%%%%%%%%%%%%%%%%%%%%%%%%%%%%%%%%%%%%%%%%%%%%%%%%%%%%%%%%%%%%%%%%%%%%%%
In this section we elaborate on the connection between the
entropy-area relation (\ref{universal}) and the microstate counting.
We consider a theory with two gauge fields in any dimension. The
statistical entropy of this theory is known and it has been verified
that the microstate entropy can be reproduced from gravity side, for
example by including Lovelock corrections with an appropriate tuning
of coefficients~\cite{Prester:2005qs}. Here we would like to change
our viewpoint to a different side. We include only general quadratic
curvature corrections and check whether the entropy-area relation
(\ref{universal}) leads to the statistical entropy. The action under
consideration in the string frame is
\begin{eqnarray}
I &=& \frac1{16\pi G} \int d^Dx \sqrt{-g} S \, \Bigl[ R + S^{-2}
(\partial S)^2 - T^{-2} (\partial T)^2 - T^2 \left( F^{(1)}_{[2]}
\right)^2 - T^{-2} \left( F^{(2)}_{[2]} \right)^2
\nonumber\\
&& \qquad + \alpha' \left( a R^2 - b R_{AB}^2 + c R_{ABCD}^2 \right)
\Bigr],
\end{eqnarray}
and the near-horizon data are exactly the ones in the previous
sections. The entropy function is
\begin{eqnarray}
f &=& \frac{\Omega_{D-2} u_S v_1 v_2^{\frac{D-2}2}}{16 \pi G}
\Biggl[ \frac{(D-2)(D-3)}{v_2} - \frac2{v_1} + \frac{2 e_1^2
u_T^2}{v_1^2} + \frac{2 e_2^2}{u_T^2 v_1^2}
\nonumber\\
&& \quad + \alpha' a \left( \frac2{v_1} - \frac{(D-2)(D-3)}{v_2}
\right)^2 - \alpha' b \left( \frac{2}{v_1^2} + \frac{(D-2)
(D-3)^2}{v_2^2} \right) + 2 \alpha' c \left( \frac2{v_1^2} +
\frac{(D-2) (D-3)}{v_2^2} \right) \Biggr].
\end{eqnarray}

The physical charges are
\begin{equation}
q_i = \frac{\partial f}{\partial e_i}, \qquad q_1=
\frac{\Omega_{D-2} u_S u_T^2 v_2^{\frac{D-2}2}}{4 \pi G v_1} \;
e_1, \quad q_2 = \frac{\Omega_{D-2} u_S v_2^{\frac{D-2}2}}{4 \pi G
u_T^2 v_1} \; e_2,
\end{equation}
and the solution for an extremal $f$ is, for $D=4$,
\begin{equation}
u_T^2 = \frac{e_2}{e_1}, \qquad v_2 = v_1, \qquad v_1 = 2 e_1 e_2 =
2 (b - 2c) \alpha', \qquad e_1 e_2 = (b - 2c) \alpha',
\end{equation}
and for $D > 4$
\begin{eqnarray}
&& u_T^2 = \frac{e_2}{e_1}, \qquad v_2 = \frac{(D-2) (D-3) v_1}2,
\qquad v_1 = \frac{4 [ 2 e_1 e_2 + \alpha' b (D-4) ]}{D^2 - 5D + 8},
\nonumber\\
&& e_1 e_2 = \frac{[ D (D-3) b - 2 (D^2 - 5D + 8) c ] \alpha'}{2
(D-2) (D-3)}.
\end{eqnarray}
Since $f=0$, the entropy and area (in the Einstein frame) are
\begin{equation}
S_{BH} = 2 \pi (e_1 q_1 + e_1 q_2) = \frac{\Omega_{D-2} u_S
v_2^{\frac{D-2}2}}{2 G v_1} (2 e_1 e_2), \qquad A = u_S \Omega_{D-2}
v_2^{\frac{D-2}2},
\end{equation}
and they are related by
\begin{equation}
S_{BH} = \frac{A}{2 G} \left( \frac{2 e_1 e_2}{v_1} \right).
\end{equation}
For $D=4$, we have $v_1 = 2 e_1 e_2$, therefore the relation $S_{BH}
= A / 2 G$ holds independently of the values of $a, b, c$ and the
value of entropy $S_{BH} = 4 \pi u_S e_1 e_2 /G$. Moreover, from the
relation~\cite{Prester:2005qs}
\begin{equation}
q_1 = \frac{2 n}{\sqrt{\alpha'}}, \qquad q_2 = \frac{2
w}{\sqrt{\alpha'}}, \qquad \frac{q_1}{q_2} = \frac{e_2}{e_1}
\frac{n}{w},
\end{equation}
we have
\begin{equation}
e_1 = \sqrt{(b - 2c) \alpha' \frac{w}{n}}, \qquad e_2 = \sqrt{(b -
2c) \alpha' \frac{n}{w}}, \qquad u_S = \frac{2 G}{\sqrt{b - 2 c} \;
\alpha'} \sqrt{nw}.
\end{equation}
The entropy can be expressed in terms of the momentum $n$ and
winding number $w$ as
\begin{equation}
S_{BH} = 8 \pi \sqrt{b-2c} \sqrt{nw}.
\end{equation}
From this check, we conclude that the entropy-area relation
(\ref{universal}) can lead to the black hole entropy to match the
statistical entropy up to a numerical factor. In order to fix this
numerical factor we should require $b - 2c = 1/4$ to match the
microstate counting $S=4\pi \sqrt{nw}$. It is interesting, as we
will see, this same additional requirement is also necessary in
any dimension.

For a general dimension, we have
\begin{equation}
e_1 = \sqrt{\alpha' \Delta \; \frac{w}{n}}, \qquad e_2=
\sqrt{\alpha' \Delta \; \frac{n}{w}}, \qquad u_S = \frac{8 \pi G
v_1}{\Omega_{D-2} v_2^{\frac{D-2}2}} \frac{\sqrt{nw}}{\sqrt{\Delta}
\; \alpha'},
\end{equation}
where
\begin{equation}
\Delta = \frac{D (D-3) b - 2 (D^2 - 5D + 8) c}{2 (D - 2) (D - 3)}.
\end{equation}
The entropy is
\begin{equation}
S_{BH} =  \frac{\Omega_{D-2} u_S v_2^{\frac{D-2}2}}{2 G v_1} (2 e_1
e_2) = 8 \pi \sqrt{\Delta} \sqrt{n w}.
\end{equation}
In order to match the microstate counting, we should have $\Delta
=1/4$. Moreover, if we first require the condition $S_{BH} = A/2G$,
then we have relation $v_1 = 2 e_1 e_2$, which gives $(D - 3) b = 2
(D -1) c$, this is identical to the result (\ref{Rbc}) for the
single charge case. But note that it is not sufficient to make
$\Delta = 1/4$. Therefore, it seems that the condition $S_{BH} =
A/2G$ and matching to microstate counting in general are two
independent requirements when we consider quadratic correction. When
$D=4$ or $5$, however, if
\begin{equation}\label{e99}
b = 4c = 1/2,
\end{equation}
these quadratic curvature terms in the action can be written in the
GB combination. In that case, we have $S=A/2G$ for both $D=4$ and
$5$. On the other hand, $\Delta =1/4$. Thus $S_{BH}=A/2G$ matches
the microstate counting. Note that $c=1/8$ from (\ref{e99}) just
gives $\lambda_2 = 1/8$ in the previous section, while the latter is
predicted by heterotic string theory.

Finally, let us note that an interesting general feature appears
once we demand the ratio of entropy and area (\ref{universal}).
Under this condition, we have $(D-3)b = 2(D-1)c$ and then $\Delta =
b - 2 c$, which is independent of spacetime dimension. Thus the
matching condition, $b - 2 c = 1/4$, yields
\begin{equation}\label{e100}
b = \frac{D-1}8, \qquad c = \frac{D-3}{16}.
\end{equation}
Although Eq.~(\ref{e100}) gives the GB combination only when $D=5$,
we expect this argument to reproduce statistical entropy can apply
to other cases. On the other hand, if we include higher-order
Lovelock terms as in \cite{Prester:2005qs}, the universal
entropy-area relation (\ref{universal}), $S_{BH}=A/2G$, can be
matched to the microstate counting, as is shown in Appendix~\ref{AppendixB}.

%%%%%%%%%%%%%%%%%%%%%%%%%%%%%%%%%%%%%%%%%%%%%%%%%%%%%%%%%%%%%%%%%%%%%%
\section{Conclusion}
\label{concl}
%%%%%%%%%%%%%%%%%%%%%%%%%%%%%%%%%%%%%%%%%%%%%%%%%%%%%%%%%%%%%%%%%%%%%%
In the Einstein gravity, the entropy of black holes is universally given
by the so-called area formula $S_{BH} = A/4G$.  However it is well
known that the area formula no longer holds in general if one
considers higher-order curvature correction terms. Still one can
calculate black hole entropy by employing the Wald's entropy formula
in the higher-order derivative gravity theories.  To calculate black
hole entropy using Wald's entropy formula, one has to know the black
hole solution. In general, however, it is difficult to find
analytical black hole solutions in higher-order derivative gravity
theories. The entropy function method proposed by Sen is a powerful
approach to get the entropy of black hole with higher-order
curvature corrections.

In this paper we showed that the entropy of small black holes with
near-horizon geometry $AdS_2\times S^{D-2}$  is always proportional
to its horizon area by employing the entropy function method and
field redefinition approach. In particular we found a universal
result that the ratio is two times of Bekenstein-Hawking
entropy-area formula in many cases of physical interest, namely
$S_{BH}=A/2G$. In four dimensions, the universal relation  always
holds irrespective of the coefficients of the higher-order terms if
the dilaton couplings are the same, which is the case for string
effective theory, while in five dimensions, the relation is again
universal irrespective of the overall coefficient if the
higher-order corrections are in the GB combination. We also
discussed how this result generalizes to known physically
interesting cases with Lovelock correction terms in various
dimensions. In the Lovelock gravity with two gauge fields, the
requirement to match the microstate counting of black holes is
consistent with the universal entropy-area relation, $S_{BH}=A/2G$.
Based on the results derived in the present paper and those in the
literature, one expects that the relation $S_{BH}=A/2G$ might be a
guidance to match the microstate counting of small black holes. Of
course, to confirm this conjecture, more evidence needs to be
accumulated.

It would also be interesting to generalize our results to more general
black hole solutions like those with deformed horizons.

\section*{Acknowledgement}
RGC and CMC are grateful to Kinki University for hospitality in
August 2007 when this work was initiated. KM and NO thank KITPC at
Beijing for hospitality during their stay when part of this work was
carried out. RGC and DWP were supported by a grant from Chinese
Academy of Sciences, grants from NSFC with No. 10325525 and No.
90403029. The work of CMC was supported by the National Science
Council under the grant NSC 96-2112-M-008-006-MY3, and in part by
the National Center for Theoretical Sciences. The work of NO was
supported in part by Grants-in-Aid for Scientific Research Fund of
the JSPS Nos. 16540250 and 06042.
The work of KM was partially supported
by  the Grant-in-Aid for Scientific Research
Fund of the JSPS (No.19540308) and for the
Japan-U.K. Research Cooperative Program,
and by the Waseda University Grants for Special Research Projects and
 for the 21st-Century
COE Program (Holistic Research and Education Center for Physics
Self-Organization Systems) at Waseda University.

\appendix

%%%%%%%%%%%%%%%%%%%%%%%%%%%%%%%%%%%%%%%%%%%%%%%%%%%%%%%%%%%%%%%%%%%%%%
\section{Area of Horizon and Field Redefinition}
\label{AppendixA}
%%%%%%%%%%%%%%%%%%%%%%%%%%%%%%%%%%%%%%%%%%%%%%%%%%%%%%%%%%%%%%%%%%%%%%
Here we derive the entropy-area relation including higher
curvature corrections from the Bekenstein-Hawking formula by a field
redefinition. Suppose we have a model with Lagrangian given by an
arbitrary function $F$ of Ricci tensor $R_{AB}$, i.e.
\begin{equation}
I = {1\over 16\pi G} \int d^Dx \sqrt{-g} \, F(g^{AB}, R_{AB}, \psi)
\,, \label{Ricci_general}
\end{equation}
where $\psi$ denotes matter field. Introducing the new metric, which is
a kind of field redefinition, as
\begin{equation}
\sqrt{-q} q^{AB} = \sqrt{-g} {\partial F \over \partial R_{AB}}
\label{Legendre} \,,
\end{equation}
we can rewrite our system as the Einstein equations with respect to
$q_{AB}$, and $g^{AB}$ behaves as a spin 2 tensor field~\cite{LT}.
The explicit form of $q^{AB}$ is not important in calculation of
black hole entropy. In this Einstein frame (we call it $q$-frame),
we have the Bekenstein-Hawking formula, that is, $S_{BH} =
A_q/4G$~\cite{KM}, where $A_q$ is the area of a black hole in
$q$-frame. Using Eq. (\ref{Legendre}), we find the entropy-area
relation in the original $g$-frame. Note that black hole entropy is
invariant under the frame transformation.

Let us discuss a concrete example. We first consider the model in
Sec.~\ref{Arbitrary} with $c=0$. In this case, we find
\begin{equation}
\sqrt{-q} q^{AB} = \sqrt{-g} \left[ \left( 1 + 2a
\mathrm{e}^{\alpha\phi} R \right) g^{AB} - 2 b
\mathrm{e}^{\alpha\phi} R^{AB} \right] \label{Legendre2} \,.
\end{equation}
Using the near-horizon solution (\ref{Ricci}) and (\ref{v1_v2}), we find
\begin{eqnarray}
\sqrt{-q} q^{\alpha\beta} &=& \sqrt{-g} \left( 1 + {2 b \over v_1}
\mathrm{e}^{\alpha \phi_s} \right) g^{\alpha\beta} \, ,
\nonumber \\
\sqrt{-q} q^{mn} &=& \sqrt{-g} \left(1 - {4 b \over (D-2) v_1}
\mathrm{e}^{\alpha \phi_s} \right) g^{mn} \label{Legendre3} \,.
\end{eqnarray}
These relations can be rewritten as
\begin{eqnarray}
q_{\alpha\beta} &=& \left( 1 + {2 b \over v_1} \mathrm{e}^{\alpha
\phi_s} \right)^{-\frac{D-4}{D-2}} \left( 1 - {4 b \over (D-2) v_1}
\mathrm{e}^{\alpha \phi_s} \right) g_{\alpha\beta} \, ,
\nonumber \\
q_{mn} &=& \left( 1 + {2 b \over v_1} \mathrm{e}^{\alpha \phi_s}
\right)^{\frac2{D-2}} g_{mn} \label{Legendre4} \,.
\end{eqnarray}
The area is given by $\sqrt{\det (g_{mn})}$ in $g$-frame and by
$\sqrt{\det (q_{mn})}$ in $q$-frame. Hence we obtain the relation
between two areas $A_q$ and $A_g$ as
\begin{eqnarray}
A_q = \left( 1 + {2 b \over v_1} \mathrm{e}^{\alpha \phi_s} \right)
A_g \,.
\end{eqnarray}
From Eqs. (\ref{v_1}) and (\ref{b}), we find
\begin{equation}
v_1 = {e^2 \over D} \mathrm{e}^{\alpha \phi_s} \, , \qquad b ={(D-2)
e^2 \over 4 D} \,.
\end{equation}
Then we have
\begin{eqnarray}
A_q = \frac{D}2 A_g \,.
\end{eqnarray}
Since we have the Bekenstein-Hawking entropy formula in $q$-frame,
we therefore obtain
\begin{eqnarray}
S_{BH} = {A_q \over 4G} = {D \over 8G} A_g \,.
\end{eqnarray}
We recover (\ref{SAanyD}) for $c=0$.

If we have Riemann tensor in the action, we cannot use this method
to calculate black hole entropy in general. However, if we restrict
our spacetime to the present metric form, i.e. $AdS_2 \times
S^{D-2}$ near horizon, we have only two metric constants $v_1$ and
$v_2$, and then we can write the Riemann curvature in terms of the
Ricci and scalar curvatures. Hence we find the effective action only
with the Ricci and scalar curvatures, which is equivalent to any
model with Lagrangian given by an arbitrary function $F$ of Riemann
tensor $R^{A}{}_{BCD}$, i.e.
\begin{equation}
I = {1 \over 16 \pi G} \int d^Dx \sqrt{-g} \, F(g^{AB},
R^{A}{}_{BCD}, \psi) \,.
\end{equation}

Here we give a simple example, which has been discussed in
Sec.~\ref{Arbitrary}. The scalar curvature $R$ and the Ricci
curvature square $R_{AB}^2$ are given by Eqs. (\ref{Ricci}) and
(\ref{Ricci_squared}). We express $v_1$ and $v_2$ in terms of $R$
and $R_{AB}^2$ as
\begin{eqnarray}
{1 \over v_1} &=& {1 \over 2D} \left[ - 2 R + \sqrt{2 (D-2) (D
R_{AB}^2 - R^2)} \right], \nonumber
\\
{1 \over v_2} &=& {1 \over D(D-2)(D-3)} \left[ (D-2) R + \sqrt{2
(D-2) (D R_{AB}^2 - R^2)} \right] \label{v1v22} \,,
\end{eqnarray}
where we have chosen the plus sign when we solve the second order
algebraic equation in order to guarantee both $v_1$ and $v_2$ are
positive.

Inserting the expression (\ref{v1v22}) into the Riemann curvature
square (\ref{Ricci_squared}), we find
\begin{equation}
R_{ABCD}^2 = {2 \over D^2(D-3)} \left[ - (D-4)^2 R^2 + D (D^2 - 5D +
8) R_{AB}^2 - 2 (D-4) R \sqrt{2 (D-2) (D R_{AB}^2 - R^2)} \right]
\,. \label{Riemann_squared}
\end{equation}
Plugging this expression into the original action (\ref{action_D}),
we obtain the equivalent action only with $R$ and $R_{AB}^2$ as
\begin{eqnarray}
I \simeq \frac1{16 \pi G} \int d^Dx \sqrt{-g} \left[ R - \frac12
(\partial \phi_s)^2 - \frac14 \mathrm{e}^{\alpha \phi_s} F^2 +
\mathrm{e}^{\alpha \phi_s} \left( \tilde{a} R^2 - \tilde{b} R_{AB}^2
- \tilde{d} R \sqrt{2 (D-2) (D R_{AB}^2 - R^2)} \right) \right] \,,
\end{eqnarray}
where
\begin{equation}
\tilde{a} = a - {2(D-4)^2 \over D^2(D-3)} c \,, \qquad \tilde{b} = b
- {2 (D^2 - 5D + 8) \over D(D-3)} c \,, \qquad \tilde{d} = {4 (D-4)
\over D^2 (D-3)} c \label{coefficients} \,.
\end{equation}
Since this action is of the form of Eq. (\ref{Ricci_general}), we
can apply the method of field redefinition.

The redefined metric (\ref{Legendre}) is now
\begin{equation}
\sqrt{-q} q^{AB} = \sqrt{-g} \left[ \left( 1 - \tilde{d}
\mathrm{e}^{\alpha \phi_s} \sqrt{2 D (D-2) R_{CD}^2} \right) g^{AB}
- 2 \tilde{b} \mathrm{e}^{\alpha \phi_s} R^{AB} \right]
\label{qg_relation} \,.
\end{equation}
Here we have used the fact that the scalar curvature $R$ vanishes
for our black hole solutions. Rewriting Eq. (\ref{qg_relation}), we
obtain
\begin{equation}
q_{mn} = \left[ 1 + 2 (\tilde{b} - D \tilde{d}) {\mathrm{e}^{\alpha
\phi_s} \over v_1} \right]^{2 \over D-2} g_{mn} \,,
\end{equation}
which implies that
\begin{equation}
A_q = \left[ 1 + 2 (\tilde{b} - D \tilde{d}) {\mathrm{e}^{\alpha
\phi_s} \over v_1} \right] A_g \label{AqAg} \,.
\end{equation}
{}From Eqs. (\ref{v_1}) and (\ref{b}), we find that
\begin{equation}
{\mathrm{e}^{\alpha \phi_s} \over v_1} = {(D-2) (D-3) \over 4 [(D-3)
b - 2c]} \,.
\end{equation}
Inserting this solution into Eq. (\ref{AqAg}) and using the
relations (\ref{coefficients}), we find
\begin{equation}
A_q = {D (D-3) b - 2(D^2 - 5D + 8) c \over 2[(D-3) b - 2c]} A_g
\label{AqAg2} \,.
\end{equation}
Note that we have $S_{BH} = A_q/4G$ in $q$-frame, and thus we
reproduce the entropy-area relation \p{SAanyD} in $g$-frame~.

In principle we can use the present method of field redefinition to
small black holes with near-horizon geometry $AdS_2\times S^{D-2}$
in any gravity theory with curvature scalar, Ricci tensor and
Riemann tensor, e.g., the Lovelock gravity discussed in the
section \ref{Lovelock}, but the equations turn out to be very complicated
in that case.

\section{Matching the microstate counting with the relation $S_{BH}=A/2G$
for two-charge system}
\label{AppendixB}
%%%%%%%%%%%%%%%%%%%%%%%%%%%%%%%%%%%%%%%%%%%%%%%%%%%%%%%%%%%%%%%%%%%%%%
In \cite{Prester:2005qs}, Prester considered small black holes in
Lovelock gravity with two gauge fields. The action is
\begin{equation} \label{a1}
I = \frac1{16\pi G} \int d^Dx \sqrt{-g} S \, \sum_{m=1}
\alpha'^{m-1} \mathcal{L}_m.
\end{equation}
where $S$ is the dilaton. The leading term in $\alpha'$ given
by~\cite{Sen:2004dp}
\begin{equation}
\mathcal{L}_1 = R + S^{-2} (\partial S)^2 - T^{-2} (\partial T)^2 -
T^2 \left( F^{(1)}_{[2]} \right)^2 - T^{-2} \left( F^{(2)}_{[2]}
\right)^2
\end{equation}
and the higher-order terms are
\begin{equation}
\mathcal{L}_m = \frac{\lambda_m}{2^m} \; \delta^{\rho_1 \sigma_1
\cdots \rho_m \sigma_m}_{\mu_1 \nu_1 \cdots \mu_m \nu_m} \; R^{\mu_1
\nu_1}{}_{\rho_1 \sigma_1} \cdots R^{\mu_m \nu_m}{}_{\rho_m
\sigma_m}, \qquad m = 2, \cdots, [D/2],
\end{equation}
where $\lambda_m$ are dimensionless parameters. By entropy function
method, it turns out that if one chooses the following set of
parameters
\begin{equation} \label{a4}
\lambda_m = \frac{4}{4^mm!},
\end{equation}
the microstate entropy $S = 4\pi \sqrt{nw}$ of the small black hole
with two charges can be reproduced by the gravity entropy of the
black hole in the action (\ref{a1}) in any dimension. Here we would
like to give a simple proof that the choice (\ref{a4}) is consistent
with the entropy-area relation (\ref{universal}), $S_{BH}=A/2G$.

With the data of the near-horizon geometry $AdS_2\times S^{D-2}$,
the entropy function is easy to calculate as \cite{Prester:2005qs}
\begin{eqnarray}
f &=& \frac{\Omega_{D-2} }{16 \pi G} u_S v_1 v_2^{(D-2)/2} \left[
\frac{2 u_T^2 e^2_1}{v_1^2} + \frac{2 e_2^2}{u_T^2 v_1^2} -
\frac{2}{v_1} - \left( \frac{1}{v_1} - \frac{2}{\alpha'} \right) A
\right]
\nonumber \\
&=& \frac{\Omega_{D-2} }{16 \pi G} u_S v_1 v_2^{(D-2)/2} g.
\end{eqnarray}
where
$$
A = A(v_2) = \sum^{[D/2]}_{m=1} \alpha'^m \lambda_{m+1}
\frac{2m(D-2)!}{(D-2m-2)!} \frac{1}{v_2^m}.
$$
Note that $\partial f/\partial u_S=0$ and $\partial f/\partial u_T =
0$ give us
\begin{equation}\label{a6}
g = 0, \qquad u_T = (e_2/e_1)^{1/2},
\end{equation}
$\partial f/\partial v_2=0$ leads to $\partial g/\partial v_2=0$,
and the latter gives
\begin{equation}\label{a7}
v_1 = \alpha' / 2.
\end{equation}
By definition, we have the physical charges $q_1$ and $q_2$ as
\begin{equation}\label{a8}
q_1 = \frac{\Omega_{D-2}}{4 \pi G} u_S v_1^{-1} v_2^{(D-2)/2} e_2,
\qquad q_2 = \frac{\Omega_{D-2}}{4 \pi G} u_S v_1^{-1} v_2^{(D-2)/2}
e_1,
\end{equation}
where we have used the relation $u_T^2 = e_2/e_1$. Furthermore,
combining (\ref{a6}) with (\ref{a7}) yields
\begin{equation}\label{a9}
e_1 e_2 = v_1/2 = \alpha'/4.
\end{equation}
Thus we can obtain the value of dilaton on the horizon through
(\ref{a8}) as
$$
u_S = \frac{8 \pi G}{\Omega_{D-2} v_2^{(D-2)/2}} \sqrt{nw}
$$
where we have used the relations, $q_1 = 2n/\sqrt{\alpha'}$ and $q_2
= 2w/\sqrt{\alpha'}$. The horizon area in the Einstein frame is
$$
A = u_S \Omega_{D-2} v_2^{(D-2)/2} = 8 \pi G \sqrt{nw},
$$
while the gravity entropy of the black hole turns out to be
\begin{equation}
S_{BH}= 2\pi (e_1 q_1 + e_2 q_2) = 4 \pi \sqrt{nw} = A/2G.
\end{equation}
Thus without knowing $v_2$, we show $S_{BH}=A/2G$. Indeed, the
parameters given in (\ref{a4}) just corresponds to the choice with
$\lambda_2=1/8$ in (\ref{lambda_m}).

\end{document}